\documentstyle[12pt]{article}
\begin{document}
\def\thefootnote{\fnsymbol{footnote}}
\begin{flushright}
KANAZAWA-94-22  \\
November, 1994
\end{flushright}
\vspace{ .7cm}
\begin{center}
{\LARGE\bf  Radiative Symmetry Breaking in a Supersymmetric Model
with an Extra $U(1)$}\\
\vspace{1 cm}
{\Large  Daijiro Suematsu}
\footnote[1]{e-mail:suematsu@hep.s.kanazawa-u.ac.jp}
\vspace{3mm} \\
{\Large  and } \vspace{3mm}\\
{\Large  Yoshio Yamagishi}
\footnote[2]{e-mail:yamagisi@hep.s.kanazawa-u.ac.jp}
\vspace {1cm}\\

{\it Department of Physics, Kanazawa University,\\
        Kanazawa 920-11, Japan}
\end{center}
\vspace{2cm}
{\Large\bf Abstract}\\
Radiative symmetry breaking is studied
in a superstring-inspired supersymmetric model which is
extended with a low energy extra
$U(1)$ symmetry.
In this model the $\mu$-problem is radiatively solved in an
automatic way.
The right-handed neutrino can be heavy and
the seesaw mechanism will produce the small neutrino mass which makes
the MSW solution applicable to the solar neutrino problem.
We search a parameter region which has the favorable feature for
the radiative symmetry breaking at the weak scale.
Rather wide parameter region is found to be allowed.
The upper bound of the extra $Z$ boson mass is estimated as
$M_{Z_2} \le 2000$~GeV for a top mass range
$ 150~{\rm GeV} \le m_t \le 190~{\rm GeV}$.
Some phenomenological features of the extra $Z$ boson are also presented.
\newpage
\setcounter{footnote}{0}
\def\thefootnote{\arabic{footnote}}
\section{Introduction}
Recently the supersymmetric theory attracts much attention.
The analyses of the precise measurements of parameters in
the standard model at LEP show that the gauge coupling
unification occurs in a precise way in the minimal supersymmetric
standard model(MSSM)\cite{lang}.
These analyses also suggest that the top quark is very heavy
($\sim 162$GeV)\cite{top}.
CDF also reports that the top quark has been found at
$m_t = 174\pm 16 $~GeV\cite{cdf}.
The heavy top quark is preferable to the radiative $SU(2) \times U(1)$
breaking in the supersymmetric model as pointed out more than ten
years ago\cite{inou}.
These facts seem to make the supersymmetric model, especially
the MSSM, more attractive than before as a particle physics model.

However, the MSSM has some unsatisfactory features.
Here we would like to stress two of them.
The first one is a hierarchy problem in relation to the
symmetry breaking, which is known as the $\mu$-problem\cite{mu}.
The MSSM has a supersymmetric Higgs mixing term $\mu \bar HH$.
To cause an appropriate radiative symmetry breaking at the weak scale
we should put $\mu \sim O(G_F^{-1/2})$ by hand, where $G_F$ is
Fermi constant.
Generally, in the supersymmetric model its typical scale is
characterized by the supersymmetry breaking scale $M_S$.
There is no reason why $\mu$ should be such a scale because it is
irrelevant to the supersymmetry breaking.
The reasonable way to answer this issue is to consider the origin of
the $\mu$ scale as a result of the supersymmetry breaking.
One of such solutions is the introduction of a singlet
field $S$ and to replace $\mu\bar HH$ by
a Yukawa type coupling $\lambda S\bar HH$.\footnote{
There are other solutions than the present one
for the $\mu$-problem\cite{kim,guid}.
However, in the models with extra gauge symmetries in the observable
sector as the $E_6$ models, other solutions do not work so easily
because
it is difficult to construct a necessary term in the gauge invariant
way.
In this paper we will not consider them.}
If $S$ gets a vacuum expectation value(VEV) of order 1~TeV
as a result of the soft supersymmetry breaking effect,
$\mu\sim O(G_F^{-1/2})$ will be realized dynamically as
$\mu =\lambda\langle S\rangle$.
A lot of works have been done on this type of models, in which the
superpotential contains the terms
$\lambda S\bar HH +\kappa S^3$\cite{mu,desa,deren}.
{}From such studies it is known that the radiative symmetry breaking can
occur successfully in a certain parameter region.

The second one is that there is no explanation for small
neutrino masses in the MSSM framework.
All known observations of the neutrino flux from the sun imply a
deficit from the value predicted by the standard solar model\cite{solar}.
It is very likely that these phenomena are explained by the new
neutrino properties.
In order to introduce the small neutrino masses which can give a
suitable solution of the solar neutrino problem,
the MSSM must be extended in a certain way.

The most consistent supersymmetric model including the gravity
is considered to be the superstring model\cite{stri}.
There is a lot of progress in this model but the N=1 supergravity
model can not be constructed as the satisfactory low energy effective
theory of superstring still now.
However, it has been shown by many efforts that its low energy gauge
structure often contains extra gauge groups other than
$SU(3) \times SU(2)
\times U(1)$, especially, the extra $U(1)$s\cite{dine,matsu}.
These extra $U(1)$ symmetries should also be broken due to some
dynamical mechanisms
because in these models the mass scale other than the Planck scale
$M_{\rm pl}$ is introduced only through the soft supersymmetry breaking.
These breaking scales of extra $U(1)$s induced by the
$SU(3)\times SU(2)\times U(1)$ singlet fields may be relevant to the
above mentioned problems.
The low energy extra $U(1)$ may be related to the dynamical
$\mu$-scale origin.
The existence of the extra $U(1)$ is also convenient for the
introduction of an intermediate scale without the large
scale supersymmetry breaking.
Such an intermediate scale may make neutrino masses small in a very simple
mechanism.
These aspects seem to make the extra $U(1)$ model very attractive as
an extension of the MSSM.

In these extra $U(1)$ models it will be an important issue to examine
the possibility of the radiative symmetry breaking of the low energy
extra $U(1)$ and also their phenomenological features.
In particular, from the phenomenological point of view it is interesting
to study the mass bound of the extra
$U(1)$ gauge boson in the recently suggested top quark mass region.
These studies will be also of benefit to the superstring model building.
In this paper we will investigate these issues on the basis of the
renormalization group equations(RGEs).

This paper is organized as follows.
In section 2 we define our model and discuss its features.
We stress in what way the extra $U(1)$s can solve the above mentioned
problems in our model.
In section 3 the scalar potential is analyzed to examine
the symmetry breaking at the weak scale by using the RGEs.
{}From such a study the various mass bounds of the physical particles are
discussed.
The section 4 is devoted to the summary.

\section{An extra $U(1)$ model}
We consider a three generation rank-six model
which is expected to derived from the superstring-inspired $E_6$ model.
There are various types of such models.
However, the model which can resolve the previously mentioned problems
of the MSSM seems to be strongly restricted.
The gauge structure of our model is $SU(3)\times SU(2)\times U(1)^3$.
There are two extra $U(1)$s.
One will be related to the $\mu$-problem and the other one will be
essential for the explanation of the small neutrino mass.
The massless field ingredients are summarized schematically as
$3\cdot {\bf 27}+ \delta\cdot (\Phi +\bar \Phi)$, where
{\bf 27} stands for a fundamental representation of $E_6$ and
it contains the full one generation structure as is well-known.
$\Phi$ and $\bar\Phi$ is some components of {\bf 27} and $\bar{\bf 27}$,
respectively.
$\delta$ is a positive integer representing the multiplicity.
Although the realization of the models with $\delta \not= 0$ is nontrivial,
the existence of such solutions is known, for example, in the Calabi-Yau
compactification\cite{dine,matsu}.
The concrete field assignments are presented in Table~1.
Here it should be noted that these field assignments are different from
those of the usually considered $E_6$ models with respect to the
Higgs doublets, the color triplets and the singlet fields.
A superpotential $W$ of this model is assumed to be expressed as
\begin{equation}
 W = h_u^{abc}Q_a\bar U_bH_c + h_d^{abc} Q_a\bar D_b\bar H_c
    + h_\nu^{abc} L_aH_bN_c + h_e^{abc} L_a\bar H_b \bar E_c
+\lambda^{abc} S_a\bar H_bH_c +k^{abc}S_ag_b\bar g_c,
\label{W}
\end{equation}
where $a, b, c(=1 \sim 3)$ are the generation indices.
As discussed later, one of the extra $U(1)$ symmetries ($U(1)_{y_E}$) is
considered to remain unbroken until the $O$(1TeV).
Since in our field assignments the singlet field $S$ has a non-zero
charge of this low
energy extra $U(1)$ which will be clarified to be  relevant to the
$\mu$-problem later, the extra isosinglet colored fields $g$ and
$\bar g$ can not be superheavy
and remain massless until the low energy region in general.
For the proton stability these extra color triplets $g$ and $\bar g$
should be assumed to decouple from the MSSM contents in the
superpotential $W$ due to some discrete symmetries\cite{dine,matsu}.
Such symmetries can be introduced and are usually expected to exist
in the superstring-inspired models.\footnote{
Here we assume the complete decoupling of $g$ and $\bar g$ in eq.(1),
for simplicity.
However, in order to prohibit the fast proton decay
it is not necessary to impose such a strong conditions.
There are many works on this problem.
Recently in ref.\cite{cm} such a possibility is discussed in some details
based on the discrete gauge anomaly cancellation.}

Now we shall point out various features related to the extra $U(1)$s of our
model.
First of all we show how the neutrinos can get the small masses.
Here we confine ourselves to the one generation case for simplicity.
The extension to the three generation case is straightforward.
As suggested in the certain superstring models\cite{dine,matsu},
two kinds of massless neutrino-like chiral superfields can exist.
One appears in pairs as (${\cal N},\bar {\cal N}$) which comes from
$(\Phi +\bar \Phi)$.
One should note that $\bar{\cal N}$ is also the chiral superfield
with the opposite charge to that of ${\cal N}$.
For the other one we use the notation $N$ to represent it.
It belongs to $3\cdot {\bf 27}$ and
appears without a complex conjugate partner.
Its fermionic component
is recognized as a right-handed neutrino.

The pair of the chiral superfields (${\cal N},\bar {\cal N}$) has
an extra $U(1)$ D-term flat direction
$\langle {\cal N} \rangle =\langle\bar {\cal N}\rangle \equiv u$
because the D-term scalar potential for (${\cal N},\bar {\cal N}$)
is proportional to  $g_{E^\prime}^2(\vert{\cal N}\vert^2-\vert\bar
{\cal N}\vert^2)^2$.
As a result, $u$ is allowed to be an intermediate
scale without breaking the supersymmetry at that scale\cite{dine}.
The existence of the extra $U(1)$ prohibits the renormalizable terms for
their self-interaction in the superpotential so that the superpotential
for them contains only the nonrenormalizable terms.
We assume that the lowest nonrenormalizable terms of the superpotential
for these fields have the following form due to a certain discrete
symmetry
\begin{equation}
W =  {c_0 \over M_{\rm pl}}\bar{\cal N}^2N^2
    + {c_n \over M_{\rm pl}^{2n-3}}({\cal N}\bar{\cal N})^n,
\label{sup}
\end{equation}
where $n$ is an integer such as $n\ge 2$\cite{lutkin}.
$c_0$ and $c_n$ are some constants.
The scale of $u$ is determined by the minimization of the scalar
potential derived from eq.(\ref{sup}),
\footnote{The supergravity correction to the scalar potential is sufficiently
suppressed by the inverse powers of $M_{\rm pl}$ so that the scalar
potential can be reduced to the global supersymmetric one
in the present case.}
\begin{equation}
V=c_n^2{u^{4n-2} \over M_{\rm pl}^{4n-6}}-M_N^2u^2.
\end{equation}
$M_N$ is the soft breaking mass
of $N$ and is assumed to be
$O(1)$~TeV.
This negative mass squared may be expected
to be induced due to the special modular weight of
${\cal N}$\cite{bribmu} and/or a radiative effect\cite{zog}.
{}From eq.(3) the minimum of $V$ is found to be realized at the intermediate
scale $u \sim (c_n^{-2}M_{\rm pl}^{4n-6}M_N^2)^{1 \over 4n-4}$.
Once these fields get such a VEV, the right-handed Majorana neutrino
mass is produced due to the first term of eq.(2) as
\begin{equation}
M_R \sim c_0(c_n^{-1}M_{\rm pl}^{n-2}M_S)^{1 \over n-1}.
\end{equation}
The mass matrix of the neutrino sector is written on the basis
$(L^0, N)$ as
\begin{equation}
{\cal M}=\left( \begin{array}{cc}
0 & h_\nu \langle H\rangle \\ h_\nu \langle H\rangle & M_R
 \end{array} \right)
\end{equation}
and the remaining neutral fermions completely decouple from these fields.
Thus this mass matrix ${\cal M}$ can present the sufficiently
small Majorana neutrino mass due to the seesaw mechanism as\cite{see}
\begin{equation}
m_\nu \sim {(h_\nu \langle H\rangle)^2 \over
c_0(c_n^{-1}M_{\rm pl}^{n-2}M_S)^{1 \over n-1}}.
\end{equation}
As a typical example, let us take $h_\nu\langle H\rangle \sim 1$~GeV as
the quark sector and put $c_0=c_n=O(1)$ and $n=3$.
For such values we get $u \sim 10^{15}$GeV and then
$M_R \sim 10^{11}$GeV.
This induces $m_\nu \sim 10^{-2}$eV,
which is suitable to the MSW solution for the solar neutrino
problem\cite{solarm}.\footnote{The similar mechanism is proposed
in refs.\cite{cve}.}
This mechanism shows that the intermediate extra $U(1)$ is
benefit for the explanation of small neutrino masses.

On the other hand a low energy extra $U(1)$ symmetry can play an
important role to solve the $\mu$-problem.
Similar model which is often called next-to-MSSM(NMSSM)
\cite{mu,desa,deren} also contains a singlet Higgs $S$.
In this type of model the $\kappa S^3$ term in the superpotential
prepares the quartic coupling for the scalar
component of $S$ and also prohibits a
massless axion.
The similar role is played by the extra $U(1)$ in the present model.
The D-term of this extra $U(1)$ supplies the
quartic coupling for the singlet $S$ and then guarantees the vacuum
stability for $S$.
The axion problem also disappears because of the existence of this
extra $U(1)$ gauge symmetry.
The detailed study of the $\mu$-problem needs the numerical analysis
of the RGEs and it will be the main subject in the next section.

We should also comment on the CP phases in the soft supersymmetry
breaking terms.
In the present model the $\mu$-term is replaced by the Yukawa type
interaction term.
Then the corresponding soft supersymmetry breaking term becomes the
ordinary A-term.
This may be thought as one of the preferable features of our model
because in such a case the physical CP phases in the soft supersymmetry
breaking terms which contribute
to the neutron electric dipole moment can be sufficiently suppressed
in an automatic way if the origin of the supersymmetry breaking
satisfies the rather suitable\cite{bribmu}.
This is based on the following mechanism that the
phase structure of the $A$-term is similar to the one of the gaugino mass
$M_a$ so that the fortunate cancellation can occur in the physical soft CP
phase arg$(AM_a^\ast)$.
However, the usual soft breaking $B$-term corresponding to the
$\mu$-term in the MSSM has no such
property and its CP phase arg$(BM_a^\ast)$ is very dangerous for the neutron
electric dipole moment.\footnote{
These features on the soft CP phases are also discussed based on the
different mechanism in ref.\cite{choi}.}

As is obvious from the previous discussion,
the largest deviation of our model from the MSSM at the low
energy region will be seen in the neutral current sector and also
in the neutrino sector.
Here let us briefly review the structure of the neutral gauge sector
containing an extra $U(1)$ and comment on the lower bound of the VEV
of $S$.
{}From the various phenomenological reason
we assume that the Higgs fields $\bar H_3, H_3$ and
$S_3$ in the third generation alone get the VEVs as
\begin{equation}
\langle \bar H_3\rangle =\left( \begin{array}{c}  \bar v \\
0 \end{array} \right) ,
\quad
\langle H_3\rangle =\left( \begin{array}{c} 0 \\
v \end{array} \right) ,
\quad
\langle S_3\rangle = x ,
\label{vev}
\end{equation}
and the VEVs of Higgs fields in the remaining generations vanish.
For simplicity, every VEV is assumed to be real and positive.
Here the VEV $x$ will be severely constrained
by the experimental results for the
neutral current.
Putting $m_Z^2={1 \over 2}(g^2_2+g^2_1)(\bar v^2 +v^2)$, the mass matrix
of the neutral gauge bosons is expressed using the charge assignments
in Table~1 as\cite{neut,elli}
\begin{equation}
m^2_Z \left( \begin{array}{cc} 1 & 2\eta\sin\theta_W \\
    2\eta\sin\theta_W &
4{(\sin\theta_W )^2 \over \bar v^2 +v^2 }({3 \over 8}\bar v^2
+{1 \over 6}v^2 + {25 \over 24}x^2 )
  \end{array} \right)
\equiv m^2_Z \left( \begin{array}{cc}
 1 & b \\ b & a \end{array} \right),
\end{equation}
where \footnote{
There is generally one more free parameter which
represents a mixing effect.
However, we shall neglect it in the present analysis
\cite{matsue}
.}
$\eta =({1 \over \sqrt 6}v^2 -
{3 \over 2\sqrt 6}\bar v^2)/(\bar v^2+v^2)$.
The mass eigenstates are defined by using a mixing angle $\theta$ as
\begin{equation}
\left( \begin{array}{c} Z_1 \\ Z_2 \end{array} \right)
= \left(\begin{array}{cc}
\cos\theta & \sin\theta \\ -\sin\theta & \cos\theta \end{array}\right)
\left(\begin{array}{c} Z \\ Z' \end{array} \right),
\end{equation}
where $\theta$ is expressed as
\begin{equation}
\tan^2\theta ={m_Z^2 -m_{Z_1}^2 \over m_{Z_2}^2 -m_Z^2}.
\end{equation}
The mass eigenvalues $m_{Z_1}^2$ and $m_{Z_2}^2$ of eq.(8) are
\begin{equation}
m_{Z_{1,(2)}}^2={1 \over 2}m_Z^2\left[(1+a){-(+)} \sqrt{(1-a)^2 +4b^2}\right].
\end{equation}
A lighter mass eigenstate $Z_1$ should be the observed $Z^0$ and
$m_{Z_1}^2 \cong m_Z^2$.
{}From eq.(11) we find that this requires $a \gg 1, b$ or
$x \gg \bar v, v$.

In order to restrict phenomenologically the value of $x$ in the
stringent way, we refer to the neutral current data and others.
For such a purpose it is convenient to draw the contours of the mass eigenvalue
$m_{Z_2}$ and the mixing angle $\theta$ in the $(\bar v/v)$-$(x/v)$ plane.
In Fig.1 we summarize such contours.
In addition, generally there can be two different definition of the Weinberg
angle as $\sin^2\theta_W\equiv {g_1^2 \over g_2^2 +g_1^2}$ and
$\sin^2\bar\theta_W \equiv 1-{m_W^2 \over m_{Z_1}^2}$,
which are exactly equivalent in the MSSM limit ($m_{Z_1}^2 \rightarrow m_Z^2$)
at tree level.
As is well-known the radiative correction shifts them differently.
The existence of the extra $U(1)$ makes them the different quantities even
at the tree level so that after carefully subtracting the radiative
correction effect $\Delta \equiv \sin^2\theta_W -\sin^2\bar\theta_W$
will be treated as the suitable measure for its existence.
As a reference, we also draw the contours of $\Delta$ in Fig.1.
By applying the experimental results to Fig.1 we may fairly
restrict the values of VEVs.
For example, the recently published constraint\cite{theta}
on the mixing angle $\vert\theta\vert <0.01$
requires that $x/v ~{^>_\sim}~ 6$ and $M_{Z_2}>550$~GeV for $v/\bar v >5$.
However, the detailed analysis is beyond the scope of the present
article and we will only comment on this point in relation to the radiative
symmetry breaking later.

\section{Scalar Potential and RGE Study}
We now study the symmetry breaking of this model at the low energy
region.
The second extra $U(1)_{y_E^\prime}$ is assumed to be broken at the
intermediate
scale $M_R$ so that we shall not consider it in the following study.
This treatment will be justified because of this extra $U(1)$ can
be decoupled not to induce any influence to the results in this section.

As is well-known, the large Yukawa couplings are essential for
the study of the radiative symmetry breaking.
Thus in addition to the usual top Yukawa coupling
$h^{333}Q_3\bar Q_3H_3$,
the largest Yukawa couplings $k^{333}S_3g_3\bar g_3$
and $\lambda^{333} S_3\bar H_3H_3$ of the extra colored fields
and the Higgs fields to the singlet $S_3$ will be important
in the analysis of the present model.
Here we assume $k^{333}>k^{3ij}$ and $\lambda^{333} >\lambda^{3ij}~~
(i,j \not=3)$.
Other terms can be neglected safely.
The relevant terms in eq.(1) to the following investigation of the
radiative symmetry breaking are
\begin{equation}
W=hQH\bar T  +\lambda S\bar HH +kSg\bar g,
\label{W2}
\end{equation}
where we abbreviated the generation indices.
The soft supersymmetry breaking terms corresponding to eq.(\ref{W2})
are
\begin{equation}
-{\cal L}_{\rm soft} =\sum_i m_i^2\vert z_i\vert^2 -\left(A_hhQH\bar T
+A_\lambda \lambda S\bar HH +A_kkSg \bar g + {\rm h.c.}\right).
\label{soft}
\end{equation}
The first terms are the mass terms of the scalar components of all chiral
superfields which are represented by $z_i$.
The remaining terms are the trelinear couplings between the corresponding
scalar components.
We also introduce the gaugino mass terms
$\displaystyle \sum_a M_a\bar\lambda_a\lambda_a$
where $a(=E,1,2,3)$ specifies the gauge group.
We do not ask the origin of these soft supersymmetry breaking
terms here.
Using eqs.(\ref{vev}),(\ref{W2}) and (\ref{soft}),
the tree-level neutral Higgs
scalar potential can be written as
\begin{eqnarray}
V_0&=&{1 \over 8}(g_2^2+g_1^2)( v^2 - \bar v^2)^2
+{1 \over 2}g_E^2({3 \over 2\sqrt 6} \bar v^2 +{1 \over \sqrt 6} v^2
-{5 \over 2\sqrt 6} x^2)^2 \nonumber \\
&+& \lambda^2 \bar v^2 x^2
   +\lambda^2 v^2 x^2
   +\lambda^2 \bar v^2 v^2  \nonumber \\
&+&m^2_{\bar H} \bar v^2 +m^2_{H} v^2 +m_S^2 x^2 \nonumber \\
&-&2A_\lambda\lambda \bar vv x,
\end{eqnarray}
where other charged fields are assumed not to have the VEVs.

Before our detailed study of this model
we comment on a phenomenological feature
which can be found without the potential minimization.
Using this potential the neutral Higgs mass matrix can be written down.
Noting that the smallest eigenvalue of the matrix is smaller
than the smallest diagonal component,
we can find the tree level upper bound
of the lightest neutral Higgs mass.
It is expressed as
\begin{equation}
m_{h^0} \le m_Z^2 \left[\cos^22\beta + {2\lambda^2 \over g_1^2}
\sin^2\theta_W\sin^22\beta + 4\sin^2\theta_W
\left({1 \over \sqrt 6}\sin^2\beta
+{3 \over 2\sqrt 6}\cos^2\beta\right)^2\right]
\label{ubhig}
\end{equation}
where $\tan\beta =v/\bar v$.
The first two terms of RHS of eq.(\ref{ubhig}) correspond
to the bound which is derived from the usually studied extend model
with a gauge singlet\cite{dree}.
It is remarkable that
the additional term raises the bound
to some extent even at the tree level.
The excess, which is due to the existence of the extra $U(1)$ gauge
symmetry, is one of the typical features of our models.
\footnote{As suggested in ref.\cite{okada}, on the argument of
the lightest Higgs mass
in the model with singlet fields
it should be noted that
the next-to-lightest Higgs might be the first observed one
in future experiments
if the lightest Higgs is singlet-dominated.}

Now we begin on the RGE study of this potential.
The potential minimization conditions for $\bar v, v $ and $x$ are
\begin{eqnarray}
 -{1 \over 2}(g_2^2+g_1^2)(\bar v^2 -v^2)\bar v
+{1 \over 2}g_E^2{3 \over 2\sqrt 6}({3 \over 2\sqrt 6}
\bar v^2+{1 \over \sqrt 6}v^2-{5 \over 2\sqrt 6}x^2)\bar v
&& \nonumber \\
+2\lambda^2\bar vv^2+2m_{\bar H}^2\bar v
-2A_\lambda\lambda vx&=&0, \nonumber \\
{1 \over 2}(g_2^2+g_1^2)(\bar v^2 -v^2)v
+{1 \over 2}g_E^2{1 \over \sqrt 6}({3 \over 2\sqrt 6}\bar v^2
+{1 \over \sqrt 6}v^2-{5 \over 2\sqrt 6}x^2)v&& \nonumber \\
+2\lambda^2\bar v^2v+2m_{H}^2v
-2A_\lambda\lambda \bar vx&=&0,  \nonumber \\
-{1 \over 2}g_E^2{5 \over 2\sqrt 6}({3 \over 2\sqrt 6}\bar v^2
+{1 \over \sqrt 6}v^2-{5 \over 2\sqrt 6}x^2)
+2m_S^2-2A_\lambda\lambda \bar vv&=&0.
\label{mini}
\end{eqnarray}
$\bar v$ and $v$ should satisfy the
constraints for the weak boson mass
\begin{equation}
m_W^2 ={1 \over 2}g_2^2(\bar v^2+v^2)\simeq (80~{\rm GeV})^2,\quad
m_{Z_1}^2 \simeq (91~{\rm GeV})^2,
\label{mz}
\end{equation}
at the potential minimum.
We also have to check whether the charged Higgs scalars have
non-vanishing VEVs.\footnote{
We must consider the possibility of the color breaking too.
We checked whether the conditions given in
ref.\cite{desa} are satisfied. However, We should note that it
has been suggested by many
authors that the conditions are neither necessary nor sufficient
in a certain cases.}
The necessary condition for the charge conservation is
\begin{equation}
m_2^2+\lambda^2 x^2 +{g_2^2 \over 4}(\bar v^2+v^2)
+{g_1^2 \over 4}(v^2-\bar v^2)+{g_E^2 \over \sqrt 6}
({3 \over 2\sqrt 6}\bar v^2+{1 \over \sqrt 6}v^2) > 0.
\label{ch}
\end{equation}
Generally $x$ is quite large as required from the neutral current
phenomena in the present models.
Therefore this condition is automatically satisfied in the most
preferable parameter region.
Under these conditions we numerically solve
these coupled equations (\ref{mini})
whose coefficients are improved by the one-loop RGEs.
For simplicity, we took the supersymmetry breaking scale $M_S$ as $m_Z$
so that our RGEs are supersymmetric in all energy range from a
unification scale $M_X$ to $m_Z$.

In order to set up the initial conditions of the RGEs
for the gauge coupling constants
we have to estimate the unification scale $M_X$.
The present model has three generations of the complete {\bf 27}s of $E_6$ as
the massless matter contents.
Therefore the one-loop $\beta$-function coefficient of the $SU(3)$ gauge
coupling is $b_3=0$.
As a result, the gauge coupling unification scale of $SU(3)$
and $SU(2)$ becomes $M_X \sim 10^{21}$GeV, which is greater than the
string scale $M_{\rm str} \sim 4\times 10^{17}$GeV\cite{kap}.
However, this discrepancy can be consistent
if we take account of the string threshold correction caused by
the string heavy modes.
As a typical example, let us consider the case of the overall modulus
$T$.
The running coupling constant taking account of the string threshold
correction can be written as\cite{dkl}
\begin{equation}
{1 \over g_a^2(\mu)}={k_a \over g_{\rm str}^2}
-{1 \over 16\pi^2}
(b_a^\prime - k_a\delta_{GS})\log[(T+T^\ast)\vert\eta(T)\vert^4]
+{1 \over 16\pi^2}b_a\log\left({M_{\rm str} \over \mu}\right)^2
\end{equation}
where $\displaystyle b_a^\prime=b_a+2\sum_iT_a(C_i)(1+n_i)$.
$\delta_{GS}$ is a duality anomaly cancellation coefficient due to
the Green-Schwarz mechanism and $\eta(T)$ is a Dedekind function.
$n_i$ is an integer called as the modular weight of a chiral superfield $C_i$.
$T_a(C_i)$ is a second order index of the field $C_i$ with respect to
 the gauge group $G_a$.
The unification condition for $SU(3)$ and $SU(2)$ is
$k_3g_3^2(M_X)=k_2g_2^2(M_X)$.
Taking the level $k_3=k_2=1$ as usual, this unification condition
becomes
\begin{equation}
\left( {M_{\rm str} \over M_X}\right)^2=
\left( (T+T^\ast)\vert\eta(T)\vert^4\right)^{b_3^\prime -b_2^\prime
\over b_3 -b_2}.
\end{equation}
Here it is remarkable that $(T+T^\ast)\vert\eta(T)\vert^4$ is less
than one for any value of $T$.
If we consider the model in which all the modular weights of the
massless matter fields are $n_i=-1$, we get $b_a^\prime =b_a$ and
$M_{\rm str}<M_X$ can always be possible as suggested in
ref.\cite{bribmu}.
In fact, in the present case $Re~T \simeq 18$ makes the values of
$M_X$ and $M_{\rm str}$ consistent.
On the basis of this argument we put
the boundary conditions of the RGEs for the gauge coupling constants
at the unification scale $M_X$ as
\begin{equation}
g_3^2(M_X)=g_2^2(M_X)= {5 \over 3}g_1^2(M_X)=
{5 \over 3}g_E^2(M_X)=g^2_{U}.
\end{equation}

The boundary conditions for Yukawa couplings and the soft supersymmetry
breaking parameters are set up at $M_{\rm str}$ in the universal way
as usual,
\begin{eqnarray}
&&h(M_{\rm str})=h_U\;,\;\lambda(M_{\rm str})=\lambda_U\;,
\;k(M_{\rm str})=k_U, \nonumber\\
&&m_Q^2(M_{\rm str})=m_{\bar U}^2(M_{\rm str})
=m_{\bar D}^2(M_{\rm str})=m_{H}^2(M_{\rm str})
=m_{\bar H}^2(M_{\rm str}) \nonumber \\
&&\qquad =m_L^2(M_{\rm str})=m_N^2(M_{\rm str})=m_{\bar E}^2(M_{\rm str})
  =m_S^2(M_{\rm str})=m_0^2, \nonumber \\
&&A_h(M_{\rm str})=A_\lambda(M_{\rm str}) = A_k(M_{\rm str})
=\nu m_0, \nonumber \\
&&M_3(M_{\rm str})=M_2(M_{\rm str})=M_1(M_{\rm str})
=M_E(M_{\rm str})=m_{1/2}.
\end{eqnarray}
These conditions for the soft supersymmetry breaking parameters are
satisfied in the certain types of superstring models.
In fact as an interesting one there is a case called as the dilaton
dominated supersymmetry breaking\cite{dil}.
The large radius limit Calabi-Yau compactification also reveals the
similar structure.
In those cases the parameter space is largely restricted as
$m_{1/2}=\sqrt 3 m_0$ and $\nu = \sqrt 3.$

Starting from these boundary values at $M_{\rm str}$ we will solve
the full RGEs.
The free parameters of our models at $M_{\rm str}$ are the soft
supersymmetry breaking parameters $m_0$, $\nu$, $m_{1/2}$ and
Yukawa couplings $h_U$, $\lambda_U$, $k_U$.
This parameter space can be somehow narrowed by imposing the additional
constraints.
To satisfy the experimental lower bound of the gluino mass,
we require the following condition on the range of gaugino
mass $m_{1/2}$ at the string scale,
\begin{equation}
40~{\rm GeV}\leq m_{1/2} \leq 200~{\rm GeV}
\end{equation}
where the upper bound is set up on the basis of the naturalness
consideration\cite{natu}.
Also we set $\nu = \sqrt 3$, for simplicity. However,
it should be noted that this setting corresponds to the
case of dilaton-dominated SUSY breaking which is favorable for
the suppression of the EDMN as mentioned before.
Using these boundary values we execute the numerical analysis
following the previously mentioned procedure.
\\
As an important input we set up the top quark mass in the range
\begin{equation}
150~{\rm GeV} \le m_t \le 190~{\rm GeV}
\end{equation}
and search the parameter space, for which the top quark mass is in this
region.
As a result, we find that the radiative symmetry breaking can occur
successfully in the rather wide parameter region.
The values of Yukawa couplings at $M_{\rm str}$ should be in the
following range,
\begin{eqnarray}
0.15 \le h_U \le 0.25,&& \nonumber \\
0.3 \le \lambda_U \le 0.9,&& \nonumber \\
0.1 \le k_U \le 0.9.&&
\end{eqnarray}
$h_U$ is confined in smaller region than other Yukawa couplings
because the mass of top quark should be in the above range.
The lower bounds of $\lambda_U$ and $k_U$ come from the requirement that
the favorable symmetry breaking at weak scale occurs.
Their upper bounds are related with the fact that $\langle S \rangle$
should not be generated at a very higher scale than $\langle S \rangle$
for the consistency of the analysis.\footnote{
For more detailed analysis of this point the study based on the
one-loop effective potential may be necessary\cite{gamba}. }

For the phenomenological purpose we reconstruct our results as the relation
between the physical particles masses, that is, the extra $Z$ boson mass and
the top quark mass.
Setting the top quark mass in the above range (25),
for the various set of parameters we plot the corresponding extra
$Z$ boson mass in Fig.2.
The comments on its qualitative features are ordered.
As easily understand, the large top mass region corresponds to the part
of the large $h_U$ in the parameter space.
Although we have not shown explicitly in figure,
we find that
the large values of $\lambda_U$ and $k_U$ have the tendency to bring the
plotted points upward.
This is based on the fact that
the large $\lambda_U$ and $k_U$ make the singlet mass $m_S^2$
small through the RGE of $m_S^2$.
This tendency of $k_U$ is more conspicuous than one of $\lambda_U$.
The lower bound of $m_{Z_2}$ appears associated with the lower
bounds of $\lambda_U$ and $k_U$.
In this analysis we require that $\vert m_Z -m_{Z_1} \vert \le 1$~GeV
where $m_{Z}^2 \equiv \frac{1}{2}(g_1^2 + g_2^2)(v_1^2+v_2^2)$.
If $m_{Z_2}$ is too small, $m_{Z_1}$ can not satisfy this condition.
Our present requirement is not so quantitative one and this lower bound
should not be taken seriously.
As mentioned in the last part in sec.2, we need more detailed analysis
by taking account of the constraints from the mixing angle and the
radiative correction effects in order to estimate the lower bound of $m_{Z_2}$.
However, we can make its rough estimation based on the experimental
result on the mixing angle $\theta<0.01$ and the value of
$1.4 < \tan\beta < 2.1 $ which is the result of the present RGE study.
In fact, we can read off the lower bound of $m_{Z_2}$ from Fig.1
as $290~{\rm GeV}< m_{Z_2} < 420$~GeV.
The upper bound of the extra $Z$ boson mass can be read off as
\begin{equation}
m_{Z_2} \le 2000~{\rm GeV}.
\label{up}
\end{equation}
This upper bound is crucially related with the upper bound of
the soft breaking parameters $m_0$ and $m_{1/2}$.
The result is not so sensitive to the values of $\nu$.
We should recognize that the bound in eq.(\ref{up}) is based on our
setting of $m_0$ and $m_{1/2}$.
Generally $m_0$ and $m_{1/2}$ increase, $m_{Z_2}$ becomes larger.
In order to show this feature
we show the $m_0$ dependence of the extra gauge boson mass $m_{Z_2}$
in Fig.3.
As shown in this figure, if the Yukawa couplings $\lambda_U$,
$k_U$ and $h_U$ are fixed, the relation
between $m_{Z_2}$ and $m_0$ is almost linear.
The termination of these lines at the certain values of $m_0$
represents that the condition (17) is no longer satisfied.
Also  $k_U$ is restricted when $\lambda_U$ and $h_U$ are fixed.
For instance, In the case presented in Fig.3,
the radiative symmetry breaking
can not occur successfully when $k_U$ exceeds 0.6.
The preferable range of $k_U$ depends on the value of $\lambda_U$.
We find that the large $\lambda_U$ pushes up the allowed range of
$k_U$ higher.
For example, if $\lambda_U=0.7$, the available range of $k_U$ is
$0.6<k_U<0.9$.
Even if we take account of the $m_0$ dependence of $m_{Z_2}$,
Fig.3 shows that the upper bound of $m_{Z_2}$ presented above
means to be reasonable from the viewpoint of the naturalness.
Anyway, this bound seems to be interesting enough that such extra $Z$
boson may be found in the future collider.

As we mentioned before, the extra color triplets $g$ and $\bar g$ can
not acquire the large mass at the sufficiently high energy scale in the
present model.
However, they can be heavy enough not to be detected directly in the
current experiment because their mass originate from the VEV of the
singlet field $S$.
\footnote{In principle, the gauge invariant coupling
${\cal N}{\bar D}g$ can be included in the superpotential.
However, If such coupling exists,
$\langle {\cal N} \rangle \ne 0$ causes an extremely large mixing
between $\bar D$ and $g$ and our model becomes unrealistic.
Therefore we must assume that such a term is forbidden
by a suitable discrete symmetry in the present model.}
{}From the present analysis of the radiative symmetry breaking we can
estimate the mass of the fermionic components of such extra color triplets.
We plot it against the extra $Z$ boson mass in Fig.4.
Because the dominant source of the extra $Z$ boson mass is also the VEV of the
singlet filed $S$, they are related linearly and the steepness depends
on the coupling $k$.
The phenomenology of these extra color triplets is one of the
interesting aspects of our model, although it is beyond the scope of the
present paper.
Their detail property will depend on the structure of the discrete
symmetry as discussed in ref.\cite{cm}.
Similar phenomena suggested there will also be expected in our model.

\section{Summary}
We examined the radiative symmetry breaking in a model with a low
energy extra $U(1)$ symmetry.
We stressed that the introduction of the extra $U(1)$s to the MSSM
has some preferable features for the explanation of the $\mu$-problem
and the neutrino mass.
It may be a very promising extension of the MSSM.
Such models are severely restricted by the theoretical and
phenomenological requirements.
In this paper we proposed a superstring inspired $E_6$ model
as such a favorable example.
We showed that in our model the radiative symmetry breaking of
$SU(2)\times U(1)$ occurs successfully in a certain parameter region.
For such a region the $\mu$-problem is solved dynamically and the top
quark mass can be situated in the region where the data obtained
recently suggest.
The extra $Z$ boson mass is also predicted in relation to the top
quark mass.
The upper bound of this extra $Z$ boson mass is $m_{Z_2} \le 2000$~GeV,
which may be encouraging for us to find it in the future collider.
This model also yields the large right-handed neutrino mass without
introducing the large scale supersymmetry breaking.
Thus it can give the small neutrino mass which is
appropriate to the MSW solution for the solar neutrino problem.
Its relation to the inflation scenario is also very interesting.
This issue will be treated elsewhere\cite{sy}.

\vspace{5mm}

\noindent
{\Large\bf Acknowledgement}\\
The authors thank S.~Matsumoto for helpful discussions.
The work of D.~S. is supported in part by a Grant-in-Aid for Scientific
Research from the Ministry of Education, Science and Culture
(\#05640337 and \#06220201).

\newpage

\newpage
{\Large\bf Table1}\\
 Particle contents in {\bf 27} of $E_6$ and
         extra $U(1)$ charge assignments.
${\cal N}$ has the same charges as $N$.

\vspace{.3cm}
\begin{center}
\begin{tabular}{c|cccc}
$\psi$   &  $(SU(3)_C, SU(2)_L)$  & $U(1)_y$ & $U(1)_{y_E}$ &
$U(1)_{y_{E^\prime}}$ \\
\hline\hline
$Q$  & $(3, 2)$ & 1/6 & $-1/2\sqrt 6$ & $\sqrt 10/12$\\
$\bar U$ &$(3^\ast , 1)$ & $-2/3$ & $-1/2\sqrt 6$ & $\sqrt 10/12$\\
$\bar D$ &$(3^\ast , 1)$ & 1/3  & $-1/\sqrt 6$ & $-\sqrt 10/6$\\
$L$ & $(1, 2)$ &$-1/2$ & $-1/\sqrt 6$ & $-\sqrt 10/6$\\
$\bar E$ &$(1, 1)$ & 1 & $-1/2\sqrt 6$ & $\sqrt 10/12$\\
$H$ &$(1, 2)$ & 1/2 & $1/\sqrt 6$ & $-\sqrt 10/6$\\
$\bar H$ &$(1, 2)$ & $-1/2$ & $3/2\sqrt 6$& $\sqrt 10/12$\\
$N$ &$(1, 1)$ & 0 & 0 & $\sqrt 10/3$\\
$S$ &$(1, 1)$ & 0 & $-5/2\sqrt 6$ & $\sqrt 10/12$\\
$g$ &$(3, 1)$ & $-1/3$ & $1/\sqrt 6$ & $-\sqrt 10/6$\\
$\bar g$ &$(3^\ast ,1)$ &1/3 & $3/2\sqrt 6$ & $\sqrt 10/12$\\
\end{tabular}
\end{center}
\newpage
{\Large\bf Figure Captions}\vspace{1cm}\\
{\Large\bf Fig.1}\\
The contours of the extra $Z$ boson mass $M_{Z_2}=300, 400, 500,
600$~GeV (solid lines),
the mixing angle $\theta=0.03, 0.02, 0.01$ (dashed-doted lines) and
$\Delta =0.015, 0.006, 0.004, 0.002$ (doted lines)
in the $(\bar v/v)$-$(x/v)$ plane.
\vspace{8mm}\\
{\Large\bf Fig.2}\\
The extra $Z$ boson mass corresponding to the
top quark mass.
Points are plotted for each values of $h_U$.
The values of $k_U$ and $\lambda_U$ are not
explicitly presented in this figure.
\vspace{8mm}\\
{\Large\bf Fig.3}\\
The dependence of the extra $Z$ boson mass on the soft supersymmetry breaking
scalar mass $m_0$ where we take $\lambda_U=0.5$ and $h_U=0.2$.
\vspace{8mm}\\
{\Large\bf Fig.4}\\
 The mass of the extra colored fermion $g, \bar g$ corresponding to the
extra $Z$ boson mass.
Points are plotted for under each values of $k_U$.
The values of $k_U$ and $\lambda_U$ are not
explicitly presented in this figure.
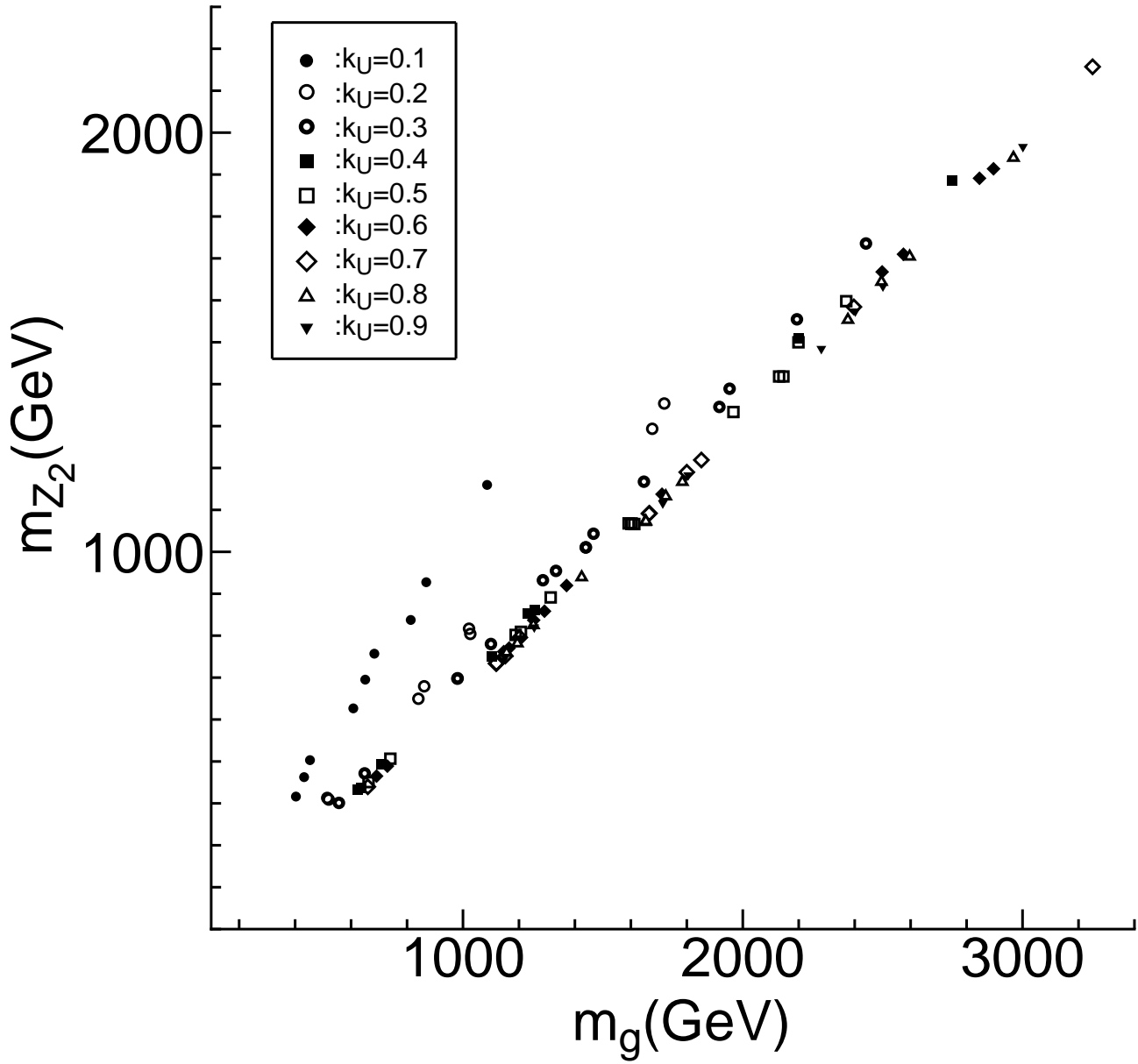
\end{document}